\documentstyle[aps,prl,twocolumn,epsfig]{revtex}
\title{Stationary states of a rotating Bose-Einstein condensate: routes to
vortex nucleation}
\author{K. W. Madison, F. Chevy, V. Bretin, and J. Dalibard}
\address{Laboratoire Kastler Brossel$^*$, D\'epartement de Physique de
l'Ecole Normale Sup\'erieure\\
24 rue Lhomond, 75005 Paris, France}
\date{submitted December 31, 2000}

\newcommand{\at}{\tilde{\alpha}}
\newcommand{\ot}{\tilde{\Omega}}
\newcommand{\eps}{\epsilon}
\newcommand{\ob}{\bar{\omega}}

\begin{document}
\maketitle 

\begin{abstract}
Using a focused laser beam we stir a $^{87}$Rb Bose-Einstein condensate 
confined in a magnetic trap.  We observe that 
the steady states of the condensate 
correspond to an elliptic cloud, stationary in the rotating
frame. These steady states depend nonlinearly on the stirring parameters
(amplitude and frequency), and various solutions can be
reached experimentally depending on the path followed in this parameter
space.  
These states can be dynamically unstable
and we observe that such instabilities 
lead to vortex nucleation in the condensate. 
\end{abstract}
{\pacs 03.75.Fi, 67.40.Db, 32.80.Lg}

\vskip 5mm

Superfluidity, originally discovered and studied in the context of 
superconductors and later in the system of superfluid liquid Helium,
 is a hallmark property of interacting quantum fluids
and encompasses a whole class of fundamental phenomena 
\cite{Lifshitz,Donnelly}. 
With the achievement of Bose-Einstein condensation 
in atomic gases \cite{Varenna99}, 
it became possible to study these phenomena in an extremely dilute 
quantum fluid, thus helping to bridge the gap between theoretical studies,
only tractable in dilute systems, and experiments.
One striking consequence of superfluidity is the response of a quantum
fluid to a rotating perturbation.  In contrast to a normal fluid, which
at thermal equilibrium will rotate like a solid body with the
perturbation, the thermodynamically stable state of 
a superfluid involves no circulation, unless the frequency of the perturbation
is larger than some critical frequency, analogous to the critical velocity
\cite{Lifshitz}.  Moreover, when the superfluid does circulate,
it can only do so by forming vortices in which the condensate density
vanishes and for which the velocity field flow evaluated around a closed
contour is quantized.

In this Letter we present the study of the response of a Bose-Einstein 
condensate (BEC) confined in a magnetic trap to a rotating perturbation
created 
by a stirring laser beam.  We observe that for
a given perturbation amplitude and frequency
the steady state of the condensate 
corresponds to an elliptic cloud, stationary in the rotating
frame as predicted in \cite{Recati00}.
Depending on the path followed in the parameter space 
(amplitude and frequency) of the rotating perturbation,
we show that two steady states can exist, corresponding to different 
ellipticities of the cloud. We also observe that 
these states possess an intrinsic dynamical instability
\cite{Recati00,CastinCom}; moreover, we find that this instability
can lead to the transformation of the elliptic state 
into a state of one or more vortices.  The fact that
vortex nucleation seems to occur only via a dynamic 
instability (apart from the phase printing method explored in
\cite{Matthews99}) explains why the frequency range over which
vortices are generated is notably smaller than that expected
from thermodynamics (\cite{Madison00} and references within).

A dilute, interacting Bose gas with a large number of atoms
is well described at low temperature by the hydrodynamic equations 
for a superfluid \cite{Stringari96}.
These equations were studied in the case of a dilute BEC in a
rotating harmonic trap characterized by the trap frequencies
$\omega_X$, $\omega_Y$, $\omega_z$ \cite{Recati00} (see also
\cite{Garcia00}).  
The axes $X,Y$ rotate
at the frequency $\Omega$ around the $z$ axis. 
There exist solutions for which
the BEC wavefunction is given by
\begin{equation}
\Psi = \sqrt{\rho} \exp\left(\frac{m}{\hbar}i \alpha X Y \right)
\ .
\label{eq:wavefunction}
\end{equation}
The condensate density for these solutions 
is a paraboloid (Thomas-Fermi profile)
\begin{equation}
\rho = \frac{\mu_{\rm e}}{g}\left[1 - 
\left(\frac{X^2}{R_X^2} + \frac{Y^2}{R_Y^2} + 
\frac{z^2}{R_z^2}\right) \right]
\ ,
\label{eq:density}
\end{equation}
where $g = 4 \pi \hbar^2 a/m$ is the collisional coupling constant set by
the
$s$-wave scattering length ($a=5.5$~nm for $^{87}$Rb in the
$|F=2,m_F=2\rangle$ ground state),
$m$ is the atomic mass, 
$\mu_{\rm e}$ is the chemical potential in the rotating frame
(the density is understood to be vanishing for $\rho\le0$).
The $XY$ ellipticity of the paraboloid is related to the parameter
$\alpha$
appearing in the phase of Eq.~(\ref{eq:wavefunction})
\begin{equation}
\alpha = \Omega \; \frac{R_X^2-R_Y^2}{R_X^2+R_Y^2}\ .
\label{eq:alpha}
\end{equation}
The value for $\alpha$ is determined by the
parameters of the rotating perturbation
according to 
\begin{equation}
\at^3 + \at(1-2\ot^2) + \eps\ot = 0
\ ,
\label{eq:alpharoot}
\end{equation}
where $\at=\alpha/\ob$, $\ot=\Omega/\ob$, 
$\ob = \sqrt{(\omega_{X}^2+\omega_{Y}^2)/2}$, 
and the trap deformation 
$\eps = (\omega_{X}^{2} - \omega_{Y}^{2})/(\omega_{X}^{2} +
\omega_{Y}^{2})$.
There are up to three possible values for
$\at$ corresponding to stationary states of the condensate
for a given pair $(\ot,\eps)$. These solutions are plotted as lines
in Figs.~1 and 2. When $\ot$ varies with a fixed $\eps$ (Fig. 1),
or when $\eps$ varies with a fixed $\ot$ (Fig. 2), the possible values 
of $\at$ are located on two branches. Branch I
is a monotonic function of $\ot$ or $\eps$, while branch II exists only
for some range of the parameters and exhibits a ``back bending", 
opening the possibility for hysteretic behavior.

The experimental study of these states relies on the realization
of a rotating harmonic potential and a measurement of the 
density profile of the condensate cloud.  
The atoms are confined in an Ioffe-Pritchard magnetic trap
which provides a static, axisymmetric 
harmonic potential of the form
$U({\bf r})=m\omega_t^2(x^2+y^2)/2 + m\omega_z^2 z^2/2$.
The condensate in this potential is cigar-shaped
with a length to diameter aspect ratio of 
$\lambda = \omega_t/ \omega_z$ which was varied between
10 and 25 ($\omega_z/2\pi=11.8 \pm 0.1$~Hz). 
The atomic cloud is stirred by a focused 500~$\mu$W laser beam
of wavelength $852$~nm and waist $w_0=20 \mu$m, whose position
is controlled using acousto-optic deflectors \cite{Madison00}
 (see also \cite{Onofrio00}). The beam
creates
an optical-dipole potential which can be approximated
by 
$m\omega_t^2(\epsilon_{X}X^2+\epsilon_{Y}Y^2)/2$.
The $XY$ axes of the optical
potential are rotated at a frequency $\Omega$ producing
a rotating harmonic trap characterized by the three trap frequencies
$\omega_{X,Y}^2 = \omega_t^2 (1+\epsilon_{X,Y})$ and
$\omega_z$.  The value of the transverse trap frequencies
is measured directly by determining the frequency domain
($\omega_Y < \Omega < \omega_X$)
in which the center of mass motion 
of the cloud becomes dynamically unstable.  This measurement
provides a determination of $\epsilon$ and
$\bar{\omega} = \sqrt{(\omega_{X}^2+\omega_{Y}^2)/2}$.

To measure the state of the condensate, characterized by
$\alpha$, we perform a field-free expansion
of the condensate for a duration of 25~ms,
followed by resonant absorption imaging technique 
along the stirring axis \cite{toosmall}.
Using the formalism presented in
\cite{Storey00}, we have checked that the value
of $\alpha$ for the rotating states 
described by Eq.~(\ref{eq:wavefunction})
changes by less than 10\,\% during the expansion for the
parameter range studied.  

The experimental procedure begins with 
the preparation of a condensate in the pure magnetic potential by
a radio-frequency (rf) forced-evaporation ramp lasting 25~s.  
Starting with $10^8$ atoms
precooled to 10~$\mu$K in an optical molasses, the atomic cloud
reaches the critical temperature $T_{\rm c}\sim$~500~nK with an atom
number of $ \sim 2.5\,10^6$.  The evaporation is continued below
$T_{\rm c}$ to a temperature of or below 100~nK 
at which point $3\;(\pm 0.7)\;10^5$ atoms are left in the condensate.
 The rf frequency is then 
raised to a value 20~kHz
above $\nu_{\rm rf}^{\rm min}$, the rf frequency which corresponds
to the bottom of the magnetic potential.  This rf drive is kept present
in order to hold the temperature approximately constant.  
At this point, the stirring laser is switched on
and the condensate is allowed to evolve in the combined magnetic and
optical
potential for a controlled duration after which we perform
the field free expansion and optical detection.

Since the condensate is first created in a static harmonic potential and
then the rotating perturbation is introduced, the final state of the
condensate
depends on its evolution in a time-dependent rotating potential
characterized by the two parameters $\eps(t)$ and $\ot(t)$. 
If the time dependencies 
are slow enough, the condensate at
every instant is in a stationary state corresponding 
to the instantaneous values of $\epsilon$ and $\ot$.
In this case, the initial state of the condensate is $\at=0$,
and as $\eps$ and $\ot$ evolve so does $\at$ evolve along
a path defined by Eq.~(\ref{eq:alpharoot}).

\begin{figure}[t]
\hskip -7mm
\epsfig{file=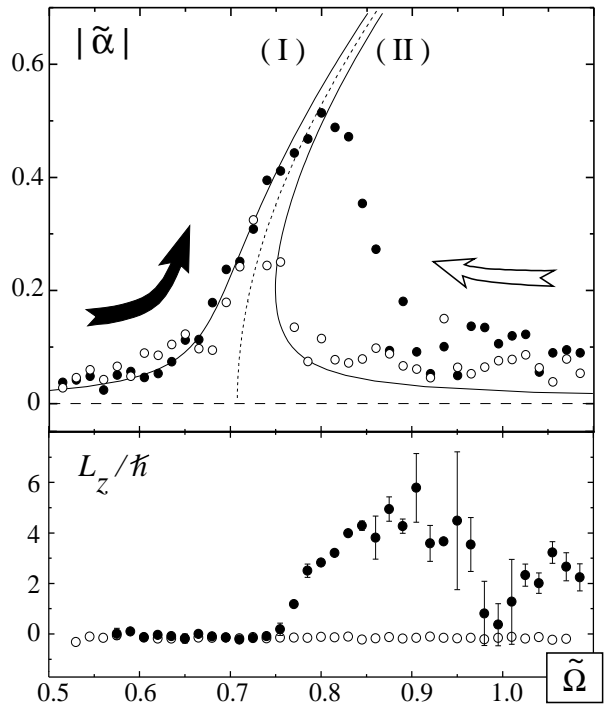,height=9.34cm,width=8.14cm}
\vskip 2mm
\caption{Steady state value of 
$\at$ as a function of stirring frequency
$\ot$.  
The results of the ascending ramp are shown by a filled dot $(\bullet)$
and of the descending ramp by a hollow dot $(\circ)$.
The branches of Eq.~(\ref{eq:alpharoot}) for 
$\eps=0$ are shown as dashed lines.
The best agreement between theory and experiment was achieved
for all data in Figs.~1 and 2 with $\eps=0.022$ and
a scaling of $\ob/2\pi=200$~Hz,
while the measured values are $\eps=0.025 \pm 0.005$
and $\ob/2\pi=195 \pm 1$~Hz.  
This 2\,\% discrepancy in the frequency 
may be due to a deviation from
the Thomas Fermi limit.
The angular momentum of the condensate measured after a
200~ms relaxation time reveals the presence of vortices
for the ascending ramp above $\ot=0.75$.}
\label{}
\end{figure}

The first study that we performed involved switching on the optical 
potential with a fixed value of $\eps=0.025 \pm 0.005$ and an initial
frequency
of $\ot=0$.  The frequency $\ot$ was increased at a constant rate  
to a final value in the range $(0.5, 1.5)$ \cite{adiabatic}.  
The rotation frequency was then held constant at this final value during
5 rotation periods, and finally the state of the condensate was measured.
For each final frequency studied, two different measurements were
performed:
a determination of $\alpha$ made according to Eq.~(\ref{eq:alpha}) and
a measurement of the angular momentum of the condensate associated with
the presence
of a vortex. The latter measurement,
which relies on the excitation of a quadrupolar oscillation
\cite{Chevy00,Haljan00}, 
is performed 200~ms after the stirring anisotropy is switched off.  This
delay
is (i) short compared to the vortex lifetime \cite{Madison00} and (ii)
long compared to the measured relaxation time ($\sim 25$~ms)
of the ellipticity of the rotating condensate.  Hence a nonzero value
of $L_z$ in this measurement is the signature of vortex nucleation.
Each measurement relies on a destructive resonant-absorption imaging, 
and is performed independently; the results are shown in Fig.~1.  
We repeated the above study
with a descending frequency ramp starting above 1
and finishing in the same range as the ascending ramp, and the results are
also
shown in Fig.~1.

\begin{figure}[t]
\hskip -5mm
\epsfig{file=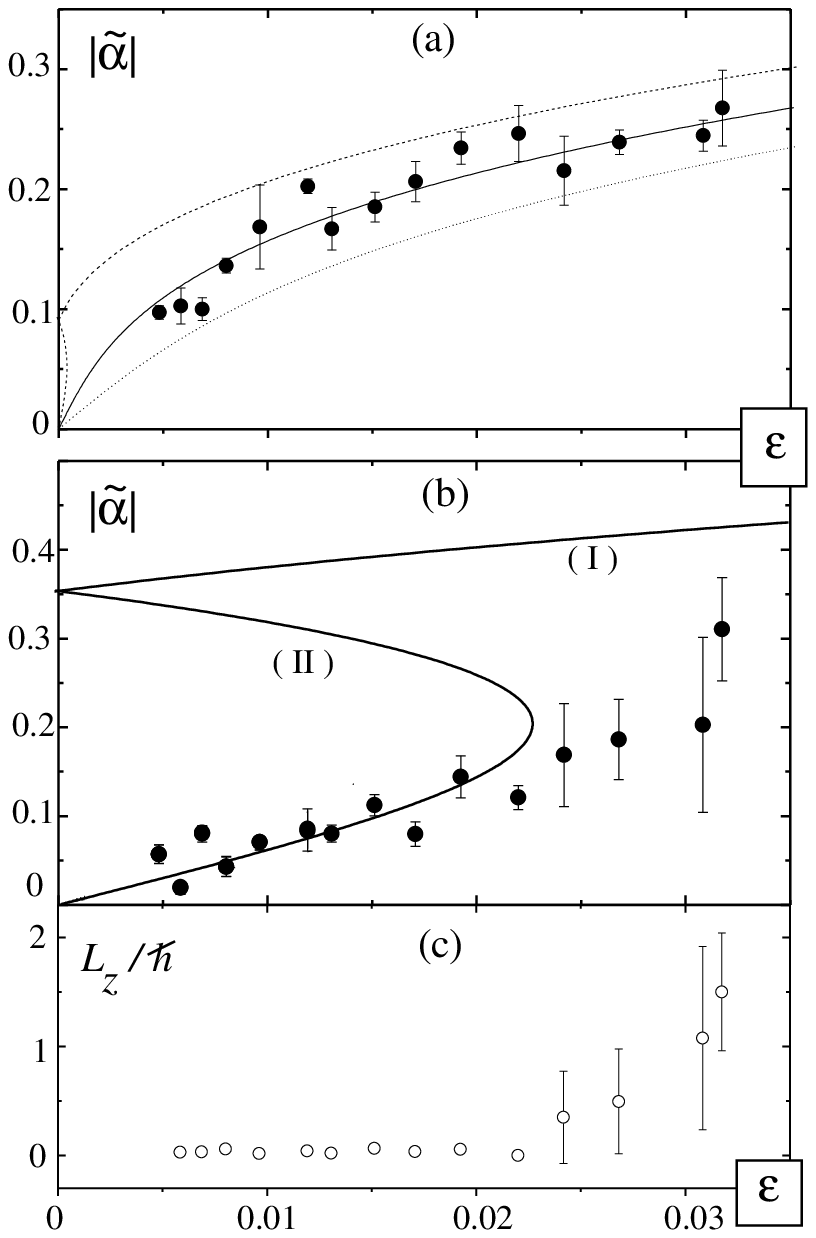,height=12.53cm,width=8.39cm}
\vskip 2mm
\caption{Measurement of the steady state value of 
$\at$ obtained from a linear ramp of the stirring anisotropy
for two different stirring frequencies.
(a) When $\ot=0.70$ the condensate follows branch I
and no vortices are nucleated.  Branch I of
Eq.~(\ref{eq:alpharoot}) is plotted for
$\ob/2\pi=200$~Hz (solid line) and also for $\ob/2\pi=197$~Hz and
$\ob/2\pi=203$~Hz (dotted and dashed lines).
(b) When $\ot=0.75$
branch II is followed to the point of back bending
($\eps_{\rm c}=0.023$).  The solid line represents
the solutions of Eq.~(\ref{eq:alpharoot}) for $\ob/2\pi=200$~Hz.
(c) For ramps which pass
this point the angular momentum measured after
a 200~ms relaxation time reveals that vortices are nucleated.}
\end{figure}

In both cases, the state of the condensate
starts on and follows one of the branches of Eq.~(\ref{eq:alpharoot}).
For the descending ramp, the lower part of branch II is followed
until the back bending frequency is reached below which this branch ceases
to be a 
solution of Eq.~\ref{eq:alpharoot}.  At this point, the condensate 
is observed to switch to branch I and to follow it down
to $\ot=0$.  The fact that no vortices are nucleated
along this path seems to confirm the prediction that
the lower part of branch II is stable \cite{Recati00}.
For the ascending ramp
the condensate is observed to start on and to follow branch I
until a frequency of $\ot_{\rm c}\simeq0.75$ is
reached.  Beyond this point the experimentally determined value of
$\alpha$ 
is no longer a solution of Eq.~(\ref{eq:alpharoot}),
and the presence of vortices is detected
in the angular momentum measurement of the condensate.
The value of $\ot_{\rm c}$ is in good agreement
with the frequency at which 
branch I is expected to become dynamically unstable \cite{CastinCom}.
This experiment was repeated for different trap geometries with
$\lambda$ between 10 and 25 and the results were unchanged.
This shows that the thermodynamic instability of the vortex state
predicted 
to occur in a very elongated geometry ($\lambda>15$) with
$\ot\sim0.75$ does not prevent vortex nucleation \cite{Feder00}.

In the second study that we present, 
the stirring frequency $\Omega$ was fixed and the
rotating deformation parameter 
$\eps$ was increased linearly from zero to a final value
in the range $(0.005 , 0.032)$ at a constant rate
$\dot{\eps}=0.078$\,s$^{-1}$.
Then, as before, we measured either $\at$ or we checked for vortex
nucleation.
In this case, since $\alpha$ is initially zero,
the branch that is followed depends on which side
of the quadrupole oscillation resonance ($\ob/\sqrt{2}$)
is $\Omega$ (see the dotted line in Fig.~1 corresponding to
$\eps=0$).  The results are shown in Fig.~2, and we observe
that when $\ot=0.7$ the condensate
follows branch I.  No vortices are nucleated in this case
which is in agreement with the predicted stability of
this branch \cite{Yvan2}.
By contrast, when $\ot=0.75$ the condensate
follows the lower part of branch II 
up to a value of $\eps_{\rm c}=0.023$ where
branch II bends backwards.  Beyond this point
the measured value of $\at$ is no longer a solution of
Eq.~(\ref{eq:alpharoot}) and we observe the nucleation of
vortices.  This situation is quite different from the one observed in
the descending frequency ramp discussed above.  In that
case, when the back bending of branch II is reached
no vortices are nucleated and the condensate is observed to
switch to branch I.  

In this second procedure, where $\ot$ is fixed and $\eps$ increases
slowly,
the critical value $\eps_{\rm c}$ at which the back bending of branch II
occurs increases with increasing $\ot$. This route can lead to vortex
nucleation
only if the final value of the ramp in $\eps$ is above $\eps_{\rm c}$,
which
for a given ramp,  
puts an upper limit on the stirring frequency leading to vortex formation.
This is well confirmed experimentally; more precisely, with our maximal
value of $\eps=0.032$, we could nucleate vortices only when the
stirring frequency was in the range $0.71 \leq \ot \leq 0.77$.

A third route to vortex nucleation is followed when the stirring
potential is switched on rapidly ($\sim20$~ms) and held constant
for 300~ms with a fixed stirring frequency \cite{Madison00,Chevy00}.  
In particular, we found that vortices are nucleated along this route
for $\ot=0.70$ and $\eps=0.032$ (final parameters in
Fig.~2a)
in marked contrast with the results of a slow ramp of $\eps$.
In this case, the state of the BEC cannot follow the sudden change
in 
$\eps$.  To illustrate the nonstationary nature of the condensate
we have plotted in Fig.~3 a typical time dependence of $\at$
with
this nucleation procedure. 
We observe that $\at$ oscillates around $0.3$ during the first
200~ms, and then falls dramatically to a fixed value below
$0.1$ at which point we see vortices entering the
condensate from the border and settling into a lattice configuration.
We should point out that the presence of the rotating anisotropy
is not necessary beyond the first $\sim80$~ms for vortex nucleation
and ordering; nonetheless, the number of vortices generated is an 
increasing function of the
stirring time after this threshold. 

\begin{figure}
\hskip -7mm
\epsfig{file=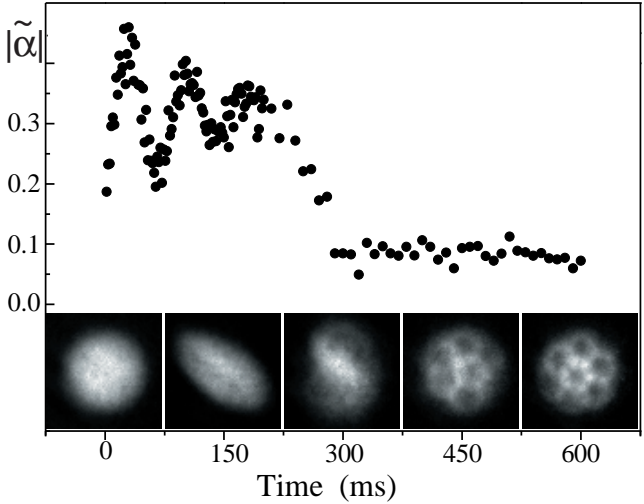,height=6.73cm,width=8.55cm}
\vskip 2mm
\caption{Measurement of the time dependence of $\at$ when the
stirring anisotropy is turned on rapidly (20~ms) to a value
of $\eps=0.025$ and a frequency of $\ot=0.7$ ($\lambda=9.2$). 
Five images taken at intervals of 150~ms show the transverse
profile of the elliptic state and reveal the 
nucleation and ordering of the resulting vortex lattice.
 The size of an image is 300$\mu$m.}
\end{figure}

In conclusion, we have studied the response of a Bose-Einstein 
condensate confined in a magnetic trap to a rotating perturbation created 
by a stirring laser beam.  We observe an irrotational state
of the condensate which corresponds to an elliptic cloud, 
stationary in the rotating frame.  We show that vortex nucleation
in this system is related to dynamical instabilities of this irrotational
state.  This connection explains why the critical frequency originally
observed in this system is notably larger than the one predicted by
a purely thermodynamic analysis: when a vortex state is
thermodynamically allowed but no dynamical instability is present, 
the time scale for vortex nucleation is probably too long for this
process to take place in our system.
A natural extension of this work is the investigation of other
irrotational
steady states associated with a rotating potential of a
higher multipole order \cite{Onofrio00}.
The exact role of temperature in vortex nucleation remains to be
elucidated.
We believe that it has conflicting roles in the generation of an ordered
vortex lattice 
(see images in Fig.~3).  On the one hand, when the temperature is
increased 
(while remaining below the condensation temperature),
the ellipticity induced by the stirring is reduced and vortex nucleation
is hindered
and may even become impossible.  On the other hand
dissipation is clearly necessary to order the vortex lattice,
and we observe qualitatively that the ordering time increases
with decreasing temperature.  We plan to address these issues in future
work.

{\acknowledgments
We thank the ENS Laser cooling
group, G. Shlyapnikov, and S. Stringari for helpful discussions. 
This work was partially supported by CNRS, Coll\`{e}ge de France,
DRET, DRED and EC (TMR network ERB FMRX-CT96-0002).  K.M.
acknowledges DEPHY for support.

$^*$ Unit\'e de Recherche de l'Ecole normale sup\'erieure et de
l'Universit\'e Pierre et Marie Curie, associ\'ee au CNRS.}


\begin{references}
\bibitem{toosmall} Since the typical radial size of the condensate
is only 3~$\mu$m, direct optical imaging is not reliable.
\bibitem{adiabatic}  The ramp lasts for 75 periods of the final frequency 
and so $d\Omega/dt$ never exceeds $2\pi$~550~Hz/s. We checked that 
at this ramp rate the condensate follows adiabatically the branches in
Fig.~1.
On the other hand the frequency ramp is fast enough so that the center of
mass 
instability does not have enough time to develop during the time that
$\ot$ spends in the instability region $[1-\eps/2,1+\eps/2]$.
\bibitem{Yvan2} Dynamical instability of this branch is expected for
$\eps>0.08$, well above the range explored experimentally 
\cite{CastinCom}.
\end{references}
\end{document}